\def\lJ{{\lambda_{_{J}}}}
\def\nL{{\mathcal{L}}}

\def\nW{{\mathcal{W}}}

\documentclass[prl,twocolumn,reprint,graphicx,showpacs,superscriptaddress,floatfix]{revtex4-1}
\usepackage{epsfig}
\usepackage{amsmath}
\usepackage{amssymb}
\usepackage{amsfonts}
\usepackage{mathptmx}
\usepackage{dcolumn}
\usepackage{eucal}
\usepackage{bm}
\usepackage{color}
\usepackage[colorlinks,linkcolor=blue,citecolor=blue]{hyperref}

\usepackage{epstopdf}

\usepackage{graphicx}

\usepackage{natbib}

\newcommand{\tmpnote}[1]%
 {\begingroup{\color{blue}\it (FIXME: #1)}\endgroup}

\begin{document}

\title{\bf Solitonic Josephson-based meminductive systems}

\author{Claudio Guarcello\thanks{e-mail: claudio.guarcello@nano.cnr.it}}
\affiliation{SPIN-CNR, Via Dodecaneso 33, 16146 Genova, Italy}
\affiliation{NEST, Istituto Nanoscienze-CNR and Scuola Normale Superiore, Piazza S. Silvestro 12, I-56127 Pisa, Italy}
\affiliation{Radiophysics Department, Lobachevsky State University, Gagarin Ave. 23, 603950 Nizhny Novgorod, Russia}
\author{Paolo Solinas\thanks{e-mail: paolo.solinas@spin.cnr.it }}
\affiliation{SPIN-CNR, Via Dodecaneso 33, 16146 Genova, Italy}
\author{Massimiliano Di Ventra\thanks{e-mail: diventra@physics.ucsd.edu}}
\affiliation{Department of Physics, University of California, San Diego, La Jolla, California 92093, USA}
\author{Francesco Giazotto\thanks{e-mail: francesco.giazotto@sns.it}}
\affiliation{NEST, Istituto Nanoscienze-CNR and Scuola Normale Superiore, Piazza S. Silvestro 12, I-56127 Pisa, Italy}

\date{\today}

\maketitle

\textbf{Memristors, memcapacitors, and meminductors, collectively called memelements, represent an innovative generation of circuit elements whose properties depend on the state and history of the system~\cite{Per11}. The hysteretic behavior of one of their constituent variables, under the effect of an external time-dependent perturbation, is their distinctive fingerprint. In turn, this feature 
	endows them with the ability to both store and process information on the same physical location, a property that is expected to benefit many applications ranging from unconventional computing to 
	adaptive electronics to robotics, to name just a few~\cite{Yan13,DiVPer13}. For all these types of applications, it is important to find appropriate memelements that combine a wide range 
	of memory states (multi-state memory), long memory retention times, and protection against unavoidable noise. Although several physical systems belong to the general class of memelements, few of them combine all of these important physical features in a single component. Here, we demonstrate theoretically a superconducting memory structure based on solitonic long Josephson junctions (LJJs). 
We show that the Josephson critical current of the junction behaves hysteretically as an external magnetic field is properly swept. 
According to the hysteretic path displayed by the critical current, a LJJ can be used as a multi-state memory, with a controllable number of available states. 
In addition, since \emph{solitons} are at the core of its operation, this system provides an intrinsic topological protection against external perturbations. 
Solitonic Josephson-based memelements may find applications as memories, and in other emerging areas such as memcomputing~\cite{DiVPer13,TraDiV15}, i.e., computing directly in/by the memory.}

Circuit elements, specifically, resistors, capacitors, and inductors with memory~\cite{Chu71,Chu76,Yan08,Str08,DiV09}, i.e., elements with characteristics that depend on the past states through which the system has evolved, have recently received increasing attention. 
Beyond the obvious applications in storing information, these elements can be combined in complex circuits to perform logic~\cite{Bor10} and unconventional computing operations~\cite{PerDiV11,Per12,Lin12,Yan13,DiV13,TraDiV15} in massive parallel schemes~\cite{DiVPer13}, 
and in the same physical location where storing occurs. Superconducting circuits that store and manipulate information are particularly appealing in view of their low-energy 
operation. Among these, a superconducting tunnel junction-based memristor was recently suggested~\cite{Peo14,Sal}. However, this type of element does not feature controllable multiple states 
that can be easily protected against unavoidable noise, due to a stochastic drift of the memory~\cite{DiV13}.  
\begin{figure*}[htbp!!]
	\centering
	\includegraphics[width=\textwidth]{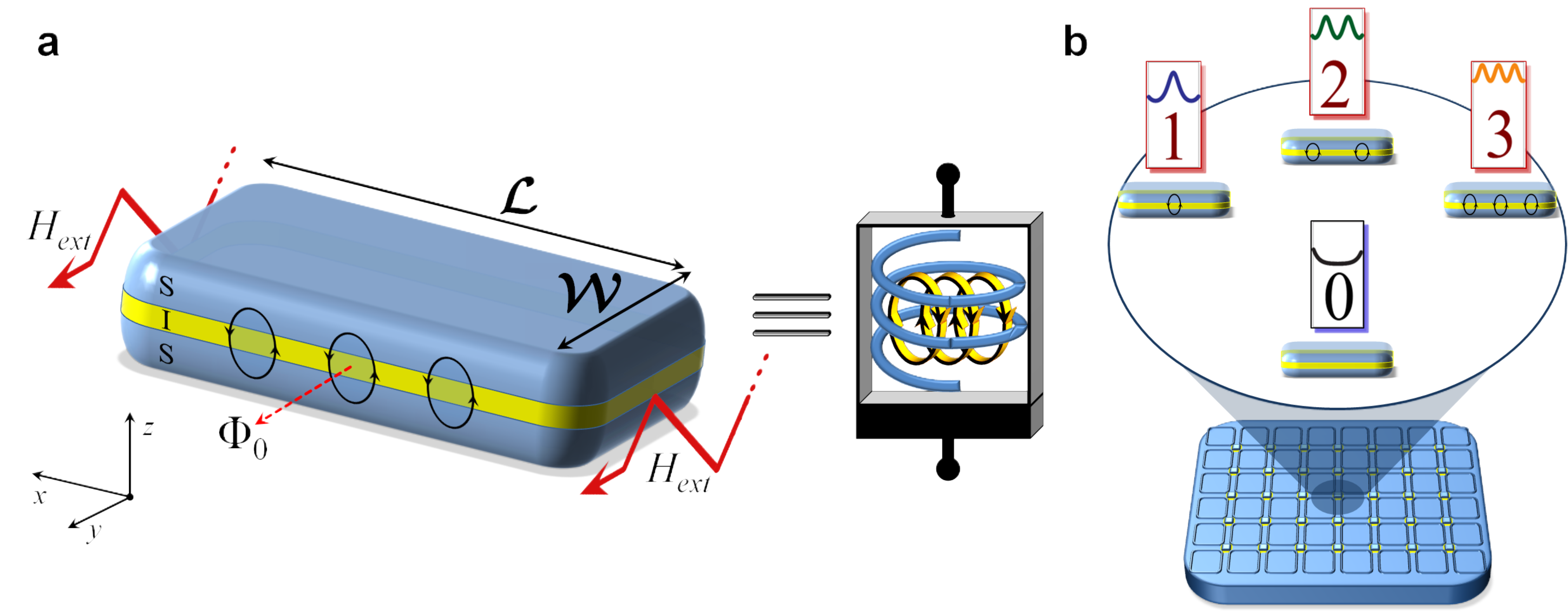}
	\caption{\textbf{Solitonic Josephson-based meminductive system.} \textbf{a}, A superconductor-insulator-superconductor (SIS) rectangular long Josephson junction (LJJ) excited by an homogeneous external periodical magnetic field $H_{ext}$. Here, we refer to the normalized field $H$ in place of $H_{ext}$ (see SI). 
		The length and the width of the junction are $\nL>\lJ$ and $\nW\ll\lJ$, respectively, where $\lJ$ is the Josephson penetration depth.
		A LJJ excited by a magnetic flux falls into the category of field-controlled meminductive systems, since the input and output variables are the applied magnetic field and the Josephson critical current, respectively. 
		The symbol used to represent the  \emph{solitonic} Josephson-based meminductive system (SJMS) is shown. 
		Fluxons ($\Phi_0$) within the junction surrounded by supercurrent loops are also represented. 
		\textbf{b}, Schematic of a possible memory drive formed by an ensemble of SJMSs. The core of the device is a LJJ excited by an in-plane magnetic field, with specific read-out electronics for the critical current. 
		As an example, we display here a junction with length $L=\nL/\lJ=10$ by which a 4-state memory element can be defined. 
		These distinct states are labelled by the number of solitons arranged along the junction. 
		The peaks in the $d\varphi/dx$ curves (see SI) and the number of loops of Josephson current surrounding the fluxons are indicated as well.}
	\label{Fig01}
\end{figure*}

Our proposal instead is based on a \emph{long} rectangular tunnel Josephson junction (LJJ) subject to a suitable periodical driving. 
A tunnel Josephson junction is a quantum device formed by sandwiching a thin insulating layer between two superconducting electrodes, and
  ``long'' refers to the physical length of the junction ($\nL$) which is supposed to exceed the Josephson penetration depth ($\lJ$).
A scheme of a LJJ with an in-plane magnetic field ($H_{ext}$) is shown in Figure~\ref{Fig01}a. 
A LJJ is the prototypical system to investigate solitons in a fully solid-state environment, and the history-dependent behavior that we envision stems from how solitons rearrange their configuration along the junction under the effect of an external magnetic field. 

\begin{figure*}[htbp!!]
\centering
 \includegraphics[width=\textwidth]{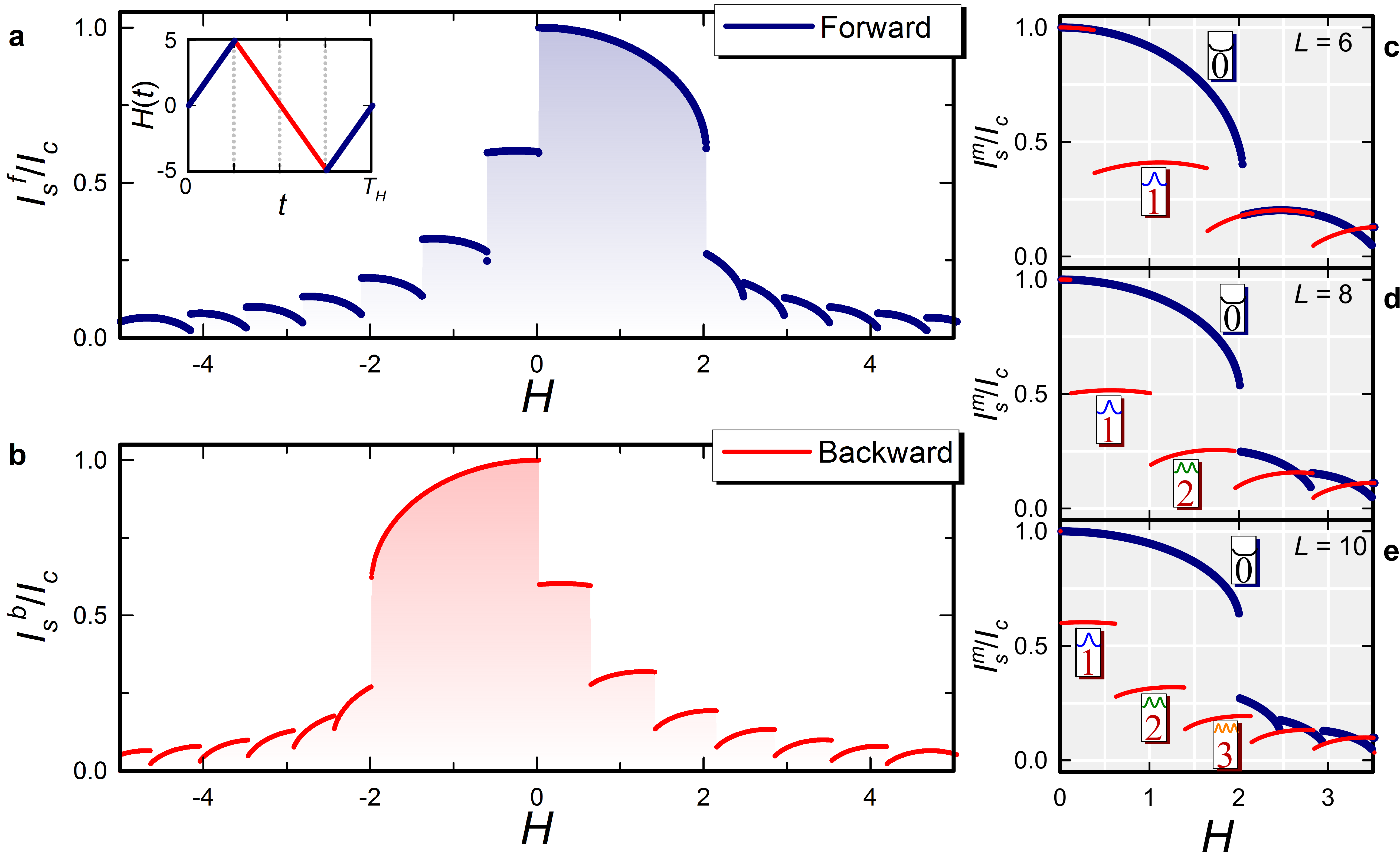}
\caption{\textbf{Josephson critical current diffraction patterns.} \textbf{a} and \textbf{b}, Normalized Josephson critical currents $I_s^f/I_c$ and $I_s^b/I_c$ as the driving field $H$ is swept forward from $H\text{=}0$ to $H\text{=}5$ (right half of panel \textbf{a}), then backward from $H\text{=}5$ to $H\text{=}-5$ (panel \textbf{b}) and again forward from $H\text{=}-5$ to $H\text{=}0$ (left half of panel \textbf{a}). 
The inset in panel \textbf{a} shows one period ($T_H$) of the driving field. The critical current as a function of $H(t)$ exhibits a diffraction-like pattern formed by lobes which are directly related to the number of solitons arranged along the junction. 
By sweeping the magnetic field forward and then backward leads to the appearance of a clear hysteretic behavior. This is a distinctive signature of any memdevice. 
According to this hysteretic behavior, the Josephson junction can be effectively used as a multi-state memory. For any specific range of magnetic field values, each state of the memory is represented by a forward or backward diffraction lobe, labeled by the number of excited solitons present along the junction. 
\textbf{c}, \textbf{d}, and \textbf{e}, Diffraction patterns for a few junction lengths $L$. 
The number of memory states provided by the SJMS can be changed by varying the junction length. The memory states associated with current lobes are indicated with the same notation as in Figure~\ref{Fig01}b.
}
\label{Fig02}
\end{figure*}

The phase dynamics of a LJJ is described by the sine-Gordon equation~\cite{Bar82,Lom82,Val14}:
\begin{equation}
\frac{\partial^2 \varphi(x,t) }{\partial t^2}+\alpha\frac{\partial \varphi(x,t) }{\partial t}-\frac{\partial^2 \varphi (x,t)}{\partial x^2} = - \sin[\varphi(x,t)].
\label{Eq01}
\end{equation}
Above, $\varphi$ is the macroscopic quantum phase difference between the superconductors, $\alpha$ denotes the intensity of the damping effect, $x$ is the spatial coordinate along the junction, and $t$ is the time (see SI).
The boundary conditions of equation~(\ref{Eq01}) read 
\begin{equation}
\frac{d\varphi(0,t) }{dx} = \frac{d\varphi(L,t) }{dx}= H(t),
\label{Eq02}
\end{equation}
where $H(t)$ is the normalized time-dependent external magnetic field, and $L=\nL/\lJ$ is the normalized length of the junction. 
By varying $H(t)$, the phase $\varphi$ evolves according to equations~(\ref{Eq01}) and~(\ref{Eq02}). 
For a spatially homogeneous supercurrent density, the Josephson critical current $I_s^m(t)$ of the junction 
shows a ``Fraunhofer-like'' diffraction pattern consisting of overlapping lobes as the magnetic field is increased, and described by the following equation~\cite{Gia13,Mar14,Gua16}:
\begin{equation}
I_s^m(t)= \frac{I_c}{L}\left|{\int_{0}^{L} dx \cos \varphi(x,t)}\right |,
\label{Eq03}
\end{equation}
where $I_c$ is the zero-field, zero-temperature junction critical current.
This behavior is shown in Figure~\ref{Fig02}a as the driving magnetic field is swept ``forward" from zero. 
A diffraction lobe corresponds to a specific number of solitons present along the junction~\cite{Gua16}. 
When the external magnetic field penetrates the junction edges it induces Josephson vortices along the weak-link, according to the nonlinearity of equation~(\ref{Eq01}). These vortices, i.e., solitons, are induced by persistent supercurrent loops carrying a quantum of magnetic flux, $\Phi_0$. The critical current, and the resulting patterns as the driving field is swept, are the physical quantities on which we focus since they can be measured with conventional techniques. In all forthcoming calculations we use parameters typical of Nb/AlOx/Nb tunnel junctions as the ideal materials combination to implement solitonic Josephson-based meminductive structures.

Figure~\ref{Fig02}b shows the diffraction pattern of the critical current when the magnetic field direction is reversed. The resulting ``backward" diffraction pattern markedly differs from the forward pattern shown in Figure~\ref{Fig02}a. For a given magnetic field $H$, the current state in which the system is found depends on the field history. This is a remarkable feature of the dissipative solitonic dynamics described by equation~(\ref{Eq01}). 
Different current states correspond to different numbers of solitons arranged along the junction, and the transition from a diffraction lobe to another corresponds to the injection, or the ejection, of solitons~\cite{Gua16}. 
As in any dissipative dynamics, the state of the system is not only determined by the value of the drive but it also depends on the path followed by the system.
This induces the forward-backward asymmetry, and the hysteretic diffraction patterns shown in Figures~\ref{Fig02}a,b.

Figures~\ref{Fig02}b-d display the forward-backward diffraction patterns as a function of the junction length.
Specifically, by increasing the length, the number of lobes forming the pattern grows, and the hysteretic asymmetry between forward and backward patterns is enhanced. 
Notably, $L$ can be tuned as well by changing the junction operation temperature ($T$) owing to the temperature dependence of $\lJ(T)$ (see SI).

The presence of both the hysteretic behavior of the critical current and highly-distinguishable current states suggests possible applications of the LJJ. For instance, this device can be used as a field-controlled memelement~\cite{Chu71,DiV09,Per11}, in which the time-dependent input/output related variables are the external magnetic field $H(t)$ and the Josephson critical current $I_s^m(t)$, respectively. 
We envisage here a memelement with distinct memory states which make use of the lobes of the forward/backward diffraction patterns. 
For a given applied magnetic field, the memelement state is determined by the value of the critical current, the latter keeping track of the field history, and pointing to a specific number of solitons present in the junction. 
Since the critical supercurrent and the magnetic field are the variables yielding the history-dependent behavior, our junction can be 
regarded as a \emph{meminductive system}~\cite{Per11,Han14}, specifically, a field-controlled solitonic Josephson-based meminductive system (SJMS). 

More generally, the LJJ can be thought as a multi-state memory in which each memory state is represented by a specific diffraction lobe, and labeled by the number of excited solitons (see Figure~\ref{Fig01}b).
For example, by referring to the diffraction patterns shown in Figures~\ref{Fig02}a,b, three backward lobes can be easily recognized within the range $H\in[0,2]$ in clear contrast to one single  forward lobe, by which a 4-state memory could be built.

\begin{figure}[t!!]
\centering
\includegraphics[width=0.50\textwidth]{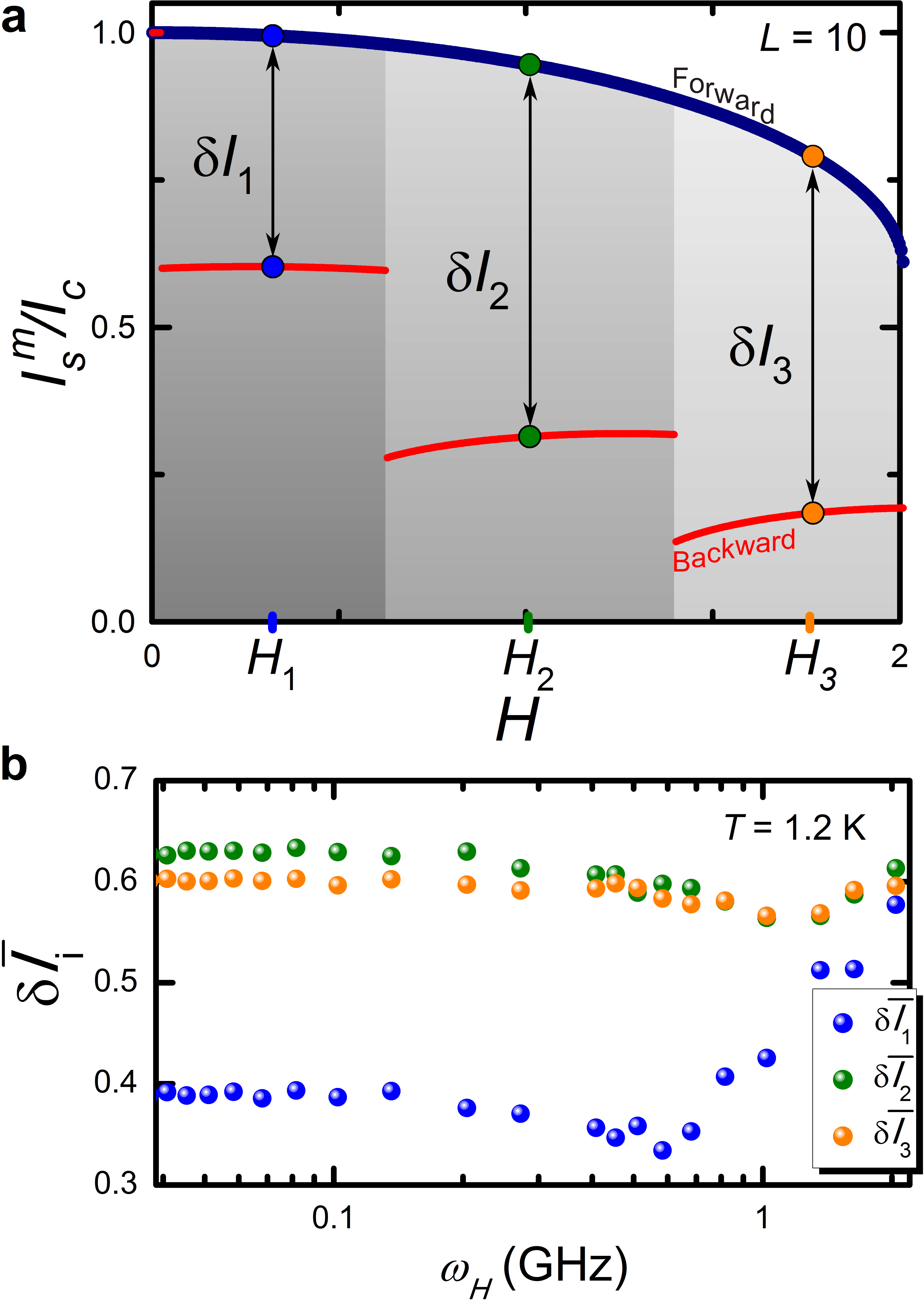}
\caption{\textbf{Frequency response of the memory states}. \textbf{a}, Forward and backward diffraction patterns for $H\in[0,2]$ and $L=10$. 
For each backward diffraction lobe, we have considered the middle magnetic field value $H_i$, and calculated the current difference $\delta I_i=\left | I^f_s(H_i)-I^b_s(H_i) \right |/I_c$  ($i=1,2,3$), where $I^f_s(H_i)$ and $I^b_s(H_i)$ are the corresponding forward and backward critical currents.
\textbf{b}, Difference $\delta \overline{I}_i$ ($i=1,2,3$) between average forward and backward diffraction patterns $\overline{I^f_s}(H_i)$ and $\overline{I^b_s}(H_i)$, computed by averaging over $N_{exp}=100$ numerical realizations of the Josephson critical current, as a function of the driving frequency $\omega_H$ for $T=1.2~$K. 
The memory states are stable up to $\omega_H\sim 0.5 \text{GHz}$. At higher frequencies, i.e.,  $\omega_H\gtrsim 1 \text{GHz}$, the system is no more able to respond to the fast driving.}
\label{Fig03}
\end{figure}
\begin{figure}[t!!]
\centering
\includegraphics[width=0.50\textwidth]{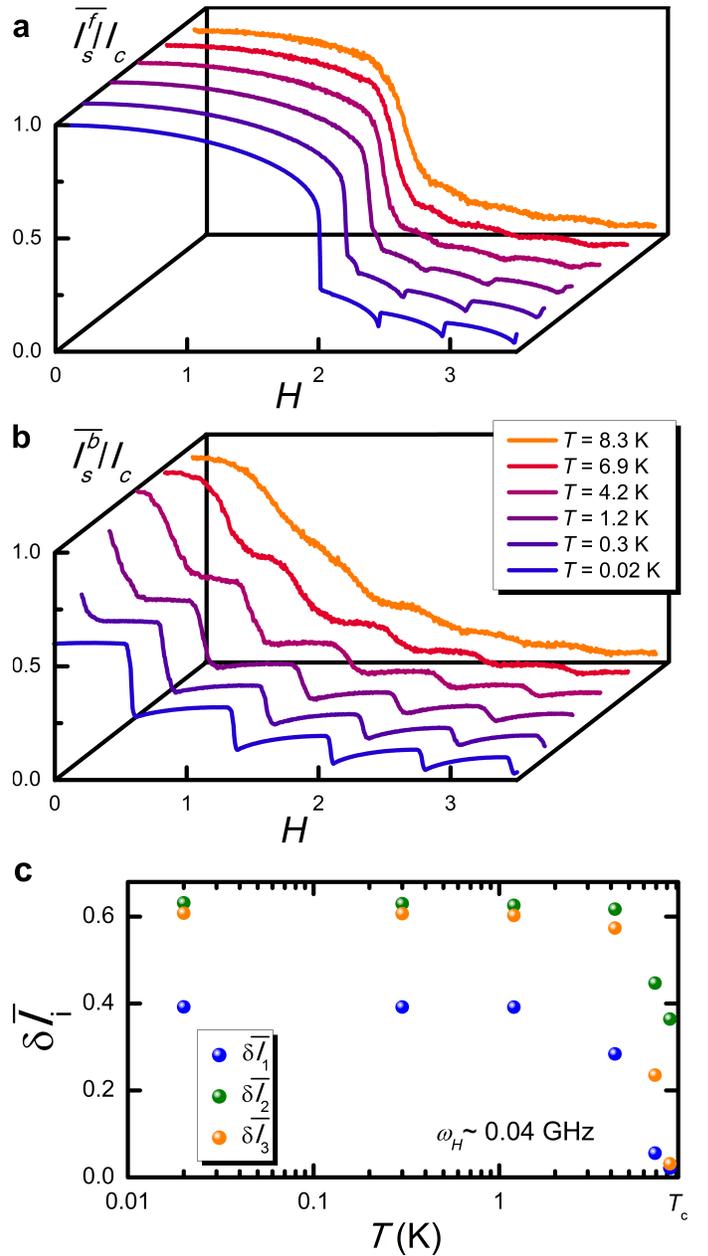}
\caption{\textbf{Effects of the temperature}. \textbf{a} and \textbf{b}, Average forward and backward diffraction patterns $\overline{I^f_s}/I_c$ and $\overline{I^b_s}/I_c$, respectively, calculated for a few temperatures, $L=10$, and $\omega_H\sim 0.04 \text{GHz}$. The patterns are computed by averaging over $N_{exp}=100$ numerical realizations of the critical current as the magnetic field is swept forward and backward when thermal fluctuations are taken into account. The legend in panel \textbf{b} refers to both panels. 
\textbf{c}, Differences $\delta \overline{I}_i$ ($i=1,2,3$) for $L=10$ and $\omega_H\sim 0.04 \text{GHz}$ calculated in correspondence of the temperatures set to obtain the results shown in panels \textbf{a} and \textbf{b}. 
By approaching the superconducting critical temperature ($T_c\simeq 9.2 \textup{K}$ for a Nb/AlOx/Nb JJ) the forward and backward diffraction patterns tend to superimpose, and $\delta \overline{I}_i$ vanishes.}
\label{Fig04}
\end{figure}

On general grounds, a good memelement has to read/write in short times, and has to be sufficiently robust against external fluctuations (noise) that tend to destroy the stored information.
On the one hand,  reading the state of the SJMS, namely the critical current $I_s^m(H)$, can be performed by conventional and well-established techniques without altering the memelement state. 
On the other hand, the writing process of each memory state depends on the operating frequency ($\omega_H$) of the magnetic field, and on the ability of the system to follow  a fast periodic driving. 
To quantify the LJJ memdevice performance as the driving frequency and the temperature are changed we make use of a figure of merit defined by
the difference between the forward and backward critical currents, $\delta I_i=\left | I^f_s(H_i)-I^b_s(H_i) \right |/I_c$, where $H_i$ is the magnetic field at the midpoint of the $i$-th backward diffraction lobe, as shown in Figure~\ref{Fig03}a for $i=1,2,3$.
For large $\delta I_i$ one can safely distinguish distinct memory states, namely, the current states. 
Furthermore, to further characterize our memdevice we have included a Gaussian thermal fluctuation term in equation~(\ref{Eq01}) (see SI) thereby making the SJMS a \emph{stochastic} memory element~\cite{Per11,Sto12,DiV13,Sli13,Pat13,Peo14} 
whereas a noiseless driving field source was considered. The relevant supercurrent differences ($\delta \overline{I}_i$) are then calculated between the averaged diffraction patterns.

Figure~\ref{Fig03}b shows $\delta\overline{I}_i$ as a function of the driving frequency $\omega_H$, for $T=1.2~$K. 
The memory states defined in Figure~\ref{Fig03}a are stable up to a driving frequency $\omega_H\sim 0.5 \text{GHz}$. At higher frequencies, i.e.,   for $\omega_H\gtrsim 1 \text{GHz}$, the system is not able to respond anymore to the fast driving.
In this region of frequencies, $\delta \overline{I}_i$ tends to increase (see SI), the diffraction patterns are not stable, and therefore cannot be used to safely distinguish the memory states.

As expected, due to its topological nature the LJJ memory shows remarkable robustness against thermal disturbances:
being a soliton-based memelement, it is intrinsically protected against small fluctuations.
Indeed, the states of the memory are associated to the number of solitons present in the LJJ and, therefore, are quantized~\cite{Gua16}.
The creation of a soliton is a macroscopic quantum phenomenon involving crossing of a potential barrier~\cite{Gua16}.
Far away from the superconducting critical temperature ($T_c$), the presence of an energy barrier in a damped dynamics prevents noise-induced state degradations, i.e., the so-called ``stochastic catastrophe''~\cite{DiV13}.

Figure~\ref{Fig04} emphasizes the robustness of the SJMS against thermal fluctuations, as the driving frequency is set to $\omega_H\sim 0.04 \text{GHz}$. 
Specifically, here we show how the temperature affects the forward (Figure~\ref{Fig04}a) and backward (Figure~\ref{Fig04}b) diffraction patterns. In particular, by increasing the temperature leads to a smoothing of the interference patterns with broadened transitions between lobes due to noise-induced creation or destruction of solitons.
Nevertheless, the memory states tend to degrade only for somewhat high temperatures approaching $T_c$ (see the results for $T>4.2~$K in Figure~\ref{Fig04}a and b). 

Finally, the stability of our Josephson-based memory as the temperature is changed is quantified in Figure~\ref{Fig04}c. 
In particular, the memory states turn out to be stable against large temperature variations, i.e., $\delta \overline{I}_i$ is roughly constant as long as $T\lesssim 4.2$~K. 
For higher temperatures, the average forward/backward diffraction patterns tend to superimpose so that  $\delta \overline{I}_i$ vanishes with the following suppression of the memory states at the critical temperature.

In summary, we have suggested long Josephson junctions excited by an external magnetic field as prototypical multi-state  superconducting memories.
Our proposal for a memory element is based on the characteristic hysteretic behavior of the critical supercurrent as the driving field is swept. 
The resulting memelement realizes a multi-state memory with a number of states controllable via the effective length of the junction. 
The solitonic nature at the origin of the critical current hysteresis makes these memory states stable and robust against thermal fluctuations. 
Our memory scheme represent the first endeavor to combine superconductivity and solitons physics in one single memelement, and could find potential application in various emerging areas such 
as logic in memory and unconventional computing~\cite{DiVPer13,TraDiV15}.

\begin{acknowledgments}

C.G. and P.S. have received funding from the European Union FP7/2007-2013 under REA
grant agreement no 630925 -- COHEAT and from MIUR-FIRB2013 -- Project Coca (Grant
No.~RBFR1379UX). 
F.G. acknowledges the European Research Council under the European Union's Seventh Framework Program (FP7/2007-2013)/ERC Grant agreement No.~615187-COMANCHE for partial financial support. M.D. 
acknowledges support from the DOE under Grant No. DE-FG02-05ER46204 and the Center for Memory and Recording Research at UCSD.
\end{acknowledgments}


\newpage

\begin{center}
{\huge Solitonic Josephson-based meminductive systems\\
\vspace{1cm}
 Supplementary Information\\
 \vspace{1cm}}
\end{center}


\section{The sine-Gordon equation and its solutions}

\begin{figure*}[t!!]
\centering
\includegraphics[width=0.57\textwidth]{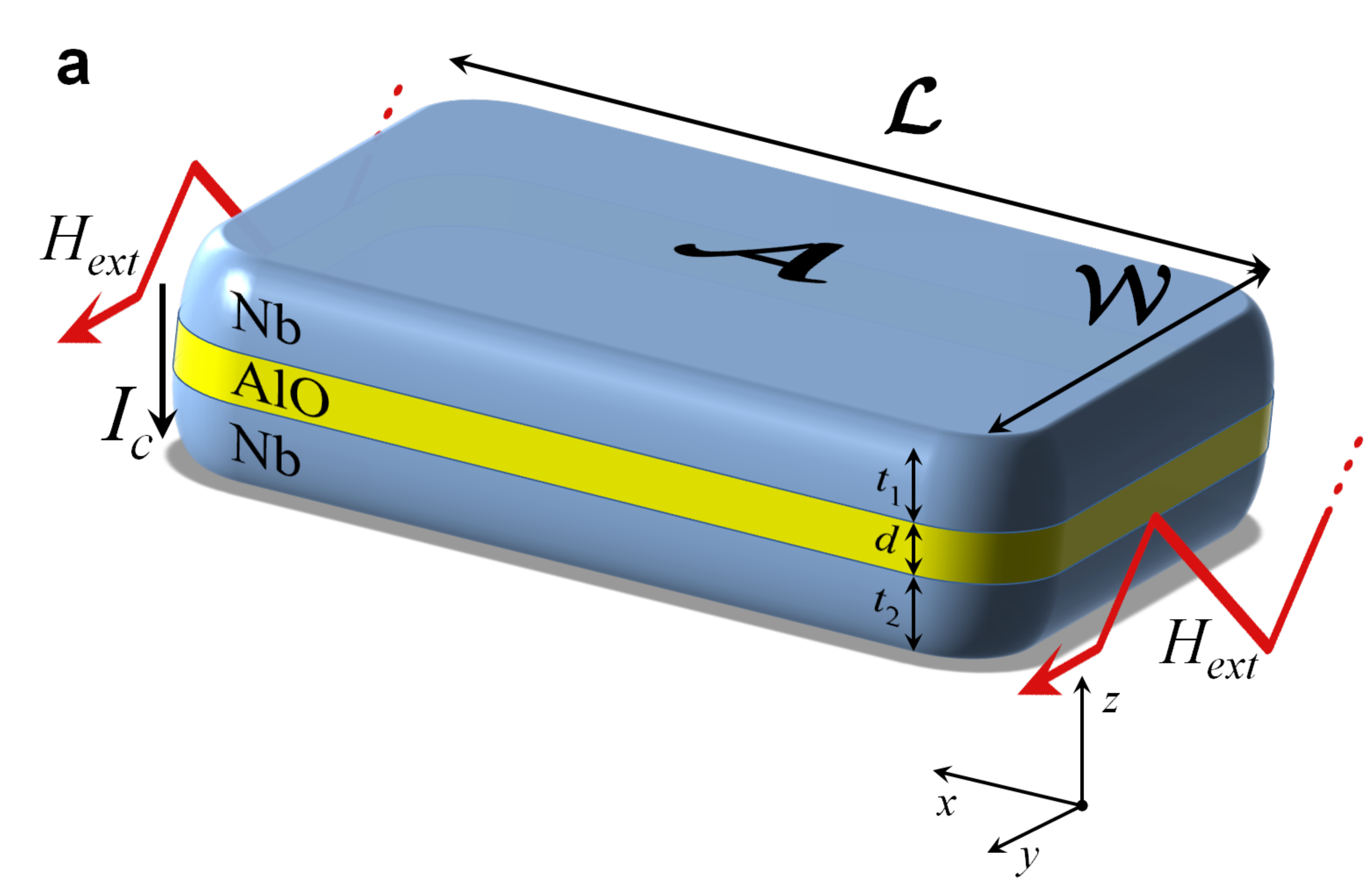}
\includegraphics[width=0.42\textwidth]{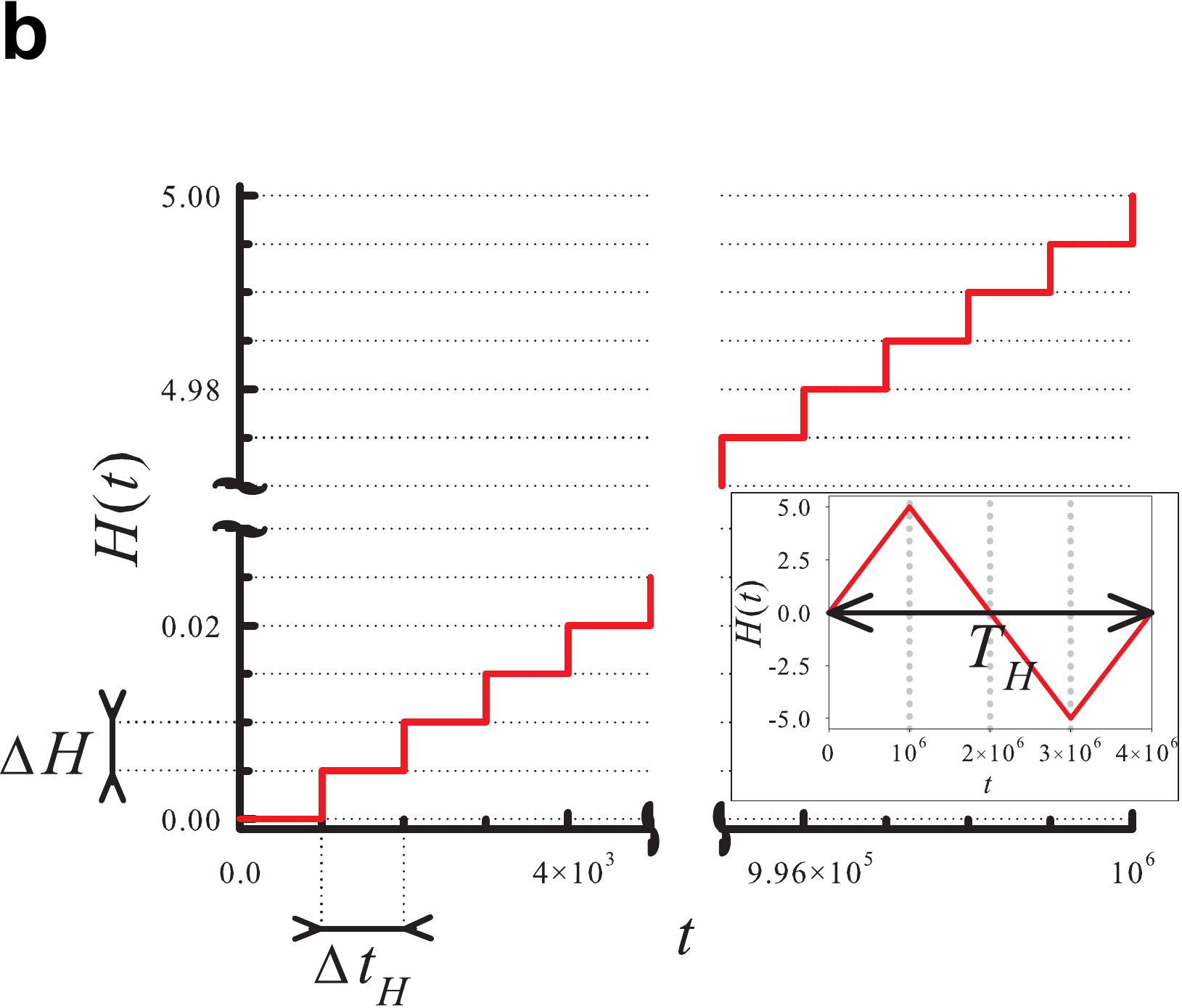}
\caption{\textbf{a}, An Nb/AlO/Nb long Josephson junction (LJJ) in the presence of a homogeneous external magnetic field $H_{ext}$ applied in the $y$ direction. The length and the width of the junction are $\mathcal{L}>\lambda_{_{J}}$ and $\mathcal{W}\ll \lambda_{_{J}}$ (according to the long junction regime), respectively, and $\mathcal{A}=\mathcal{L}\mathcal{W}$ is the junction area, $\lambda_{_{J}}$ being the Josephson penetration depth. Moreover, $t_i$ and $d$ denote the thicknesses of the \emph{i}-th superconductor and the insulating interlayer, respectively. \textbf{b}, Numerical implementation of the magnetic field drive.
Normalized staircase external magnetic field $H(t)$, formed by small steps with height $\Delta H=0.005$ kept constant for time intervals $\Delta t_H=10^3$, with $H_{max}=5$. In the inset, a driving period $T_H=4\times 10^6$ of $H(t)$ is shown. The times are normalized with respect to the inverse of the Josephson plasma frequency.}
\label{FigSI01}
\end{figure*}

In Fig.~\ref{FigSI01}(a), a long and narrow Nb/AlO/Nb Josephson junction (JJ) is represented. The electrodynamics of a long JJ (LJJ) is usually described by a partial differential equation for the order parameter $\varphi$, namely, the phase difference between the wavefunctions describing the carriers in the superconducting electrodes. In normalized units, the perturbed sine-Gordon (SG) equation reads~\cite{SILom82,SIBar82,SIVal14,SIGuaValSpa16,SIGua16} 
\begin{equation}
\frac{\partial^2 \varphi }{\partial t^2}+\alpha\frac{\partial \varphi }{\partial t}-\frac{\partial^2 \varphi }{\partial x^2} = - \sin(\varphi),
\label{SGeq}
\end{equation}
with boundary conditions taking into account the normalized external magnetic field $H(t)$
\begin{equation}
\frac{d\varphi(0,t) }{dx} = \frac{d\varphi(L,t) }{dx}= H(t).
\label{bcSGeq}
\end{equation}
In equation~(\ref{SGeq}), space and time variables are normalized to the \emph{Josephson penetration depth} $\lambda_{_{J}}$ and the inverse of the \emph{Josephson plasma frequency} $\omega_p$, respectively. They read 
\begin{eqnarray}
\lambda_{_{J}}&=&\sqrt{\frac{\Phi_0}{2\pi \mu_0}\frac{1}{t_d J_c}}\\
\omega_p&=&\sqrt{\frac{2\pi}{\Phi_0}\frac{I_c}{C}}, 
\end{eqnarray}
where $R$ and $C$ are the total resistance and capacitance of the JJ, $\Phi_0= h/2e\simeq2.067\times10^{-15} \textup{Wb}$ is the magnetic flux quantum, $\mu_0$ is the vacuum permeability, $I_c$ and $J_c=I_c/\mathcal{A}$ are the critical current and the critical current area density ($\mathcal{A}$ being the junction area). Moreover, $ t_d=\lambda_1+\lambda_2+d$
%
%
 is the effective magnetic thickness, $\lambda_i$ being the London penetration depth of the superconductor $S_i$ and $d$ the interlayer thickness. If $\lambda_i$ exceeds the thickness $t_i$ of the \emph{i}-th superconductor, the effective magnetic thickness has to be replaced by $\tilde{t}_d=\lambda_1\tanh\left ( t_1/2\lambda_1 \right )+\lambda_2\tanh\left ( t_2/2\lambda_2 \right )+d$.

The Josephson penetration depth represents the length scale of the system, so that a JJ is regarded as long and narrow if the length and the width of the junction are $\mathcal{L}>\lambda_{_{J}}$ and $\mathcal{W}\ll \lambda_{_{J}}$, respectively. In normalized unit, the linear dimensions of the junction read $L=\mathcal{L}/\lambda_{_{J}}>1$ and $W=\mathcal{W}/\lambda_{_{J}}\ll 1$. Moreover in equation~(\ref{SGeq}), $\alpha=(\omega_p RC)^{-1}$ is the damping parameter.

The SG equation admits traveling wave solutions, called \emph{solitons}~\cite{SIUst98}. In the SG framework, a soliton is often referred to as a kink.
For the unperturbed SG equation, i.e., $\alpha=0$ in equation~(\ref{SGeq}), solitons have the simple analytical expression~\cite{SIBar82}
\begin{equation}
\varphi(x-ut)=4\arctan \left \{ \exp \left [ \pm \frac{\left(x-ut \right )}{\sqrt{1-u^2}} \right ] \right \},
\label{SGkink}
\end{equation}
where the sign $\pm$ is the polarity of the soliton (specifically, the minus sign defines an \emph{antisoliton}) and $u$ is the Swihart’s velocity~\cite{SIBar82}, namely, the largest group propagation velocity of the linear electromagnetic waves in long junctions. Specifically, the phase of a soliton (antisoliton) twists from 0 to 2$\pi$ (from 2$\pi$ to 0). Alternatively, $\varphi/2\pi$ has a “topological charge” +1 for each soliton and -1 for each antisoliton.
Moreover, a SG soliton has a well defined physical meaning in the LJJ framework, since it carries a quantum of magnetic flux $\Phi_0$, induced by a supercurrent loop surrounding it, with the local magnetic field perpendicularly oriented with respect to the junction length~\cite{SIMcL82}. Thus, a soliton is usually referred to as a \emph{fluxon}, or a Josephson vortex, in the context of LJJs.

In equation~(\ref{bcSGeq}), the normalized external magnetic field is $H(t)=\frac{2\pi\mu _0}{\Phi _0}t_d\lambda_J H_{ext}(t)$, where $H_{ext}(t)$ is the non-normalized external magnetic field lying parallel to a symmetry axes of the junction and along $y$, see Fig.~\ref{FigSI01}a.
For the numerical simulation, we have modeled $H(t)$ as a staircase function formed by steps with ``treads'' deep $\Delta t_H$ and ``risers'' high $\Delta H$ [see Fig.~\ref{FigSI01}(b)] and is ramped up from zero to $H_{max}$, then reduced to $-H_{max}$ and subsequently raised again to zero to perform a double-swept drive. Accordingly, the driving period $T_H=4(H_{max}/\Delta H)\Delta t_H$ and the frequency $\omega^*_H=1/T_H$ are defined.

\section{The critical current diffraction patterns}

The $\varphi$-dependent supercurrent as a function of the external magnetic field $H$ can be expressed as
\begin{equation}
I_s ( H )=\iint dx dy J_s(x,y) =\iint dx dy J_c(x,y)\sin [\varphi(x,y)] 
\label{PhiJosephsonCurrent}
\end{equation}
where $J_s(x,y)$ is the supercurrent density per unit area, and $J_c(x,y)$ is the Josephson critical current density. We denote with $i_c(x)$ the $J_c(x,y)$ integral in the direction of the magnetic field
\begin{equation}
i_c(x)=\int_{-\mathcal{W}/2}^{\mathcal{W}/2}J_c(x,y)dy,
\label{i_c_W}
\end{equation}
so that the Josephson current becomes
\begin{equation}
I_s(H)=\int_{-\mathcal{L}/2}^{\mathcal{L}/2} i_c(x)\sin\varphi(x)dx.
\label{GenericIc}
\end{equation}
In equation~(\ref{GenericIc}), $\varphi(x)$ is the phase difference induced by the applied magnetic field $H_{ext}$. In fact, $\varphi$ depends on the local magnetic field $H_y(x)$ through the equations~\cite{SIBar82}
\begin{equation}
\frac{\partial \varphi }{\partial x}=\frac{2\pi\mu_0t_d}{\Phi_0}H_y(x)=h_y(x) \qquad \qquad \frac{\partial \varphi }{\partial y}=0.
\label{LocalMagneticField}
\end{equation}
The latter equation comes from the condition $\mathcal{W}\ll \lambda_{_{J}}$, so that $\varphi \left ( x,y \right )\equiv \varphi \left ( x \right )$.

For a \emph{short} rectangular JJ ( $\mathcal{L}\ll\lambda_{_{J}}$ and $\mathcal{W}\ll \lambda_{_{J}}$) the external magnetic field fully penetrates the junction and is spatially homogeneous along it, namely, $H_y(x)\equiv H_{ext}$, so that, according to equation~(\ref{LocalMagneticField}), the phase is just linearly increasing in the $x$ direction, 
\begin{eqnarray}
\varphi(x)=\left ( \frac{2\pi\mu_0t_d}{\Phi_0} H_{ext} \right ) x+\varphi_0=kx+\varphi_0.
\label{ShortJJIc}
\end{eqnarray}
Accordingly, equation~(\ref{GenericIc}) becomes
\begin{eqnarray}\nonumber
I_s(H)&=& \underset{-\mathcal{L}/2}{\overset{\mathcal{L}/2}{\mathop \int }} i_c(x)\sin(kx+\varphi_0)dx=\Im\left \{ \underset{-\infty}{\overset{\infty}{\mathop \int }}i_c(x)e^{i(\varphi_0+kx)}dx\right \}\\
&=&\textup{Im}\left \{ e^{i\varphi_0} \int_{-\infty }^{\infty }i_c(x)e^{ikx}dx\right \}.
\end{eqnarray}
The Josephson critical current is the amplitude of the last integral, that is
\begin{equation}
I_s^m(H)=\left |\int_{-\infty }^{\infty }i_c(x)e^{ikx}dx \right |,
\label{AbsGenericIc}
\end{equation}
independent of any phase factor $\varphi_0$.

By assuming a uniform supercurrent area density within the junction, i.e., $J_c(x,y) \equiv J_c$, for $0\leq x \leq \mathcal{L}$ and $0\leq y \leq \mathcal{W}$, and zero elsewhere, we obtain $i_c(x)\equiv i_c=J_c\,\mathcal{W}$, according to equation~(\ref{i_c_W}). Therefore, equation~(\ref{AbsGenericIc}) becomes
\begin{eqnarray}\nonumber
I_s^m(H)&=&\left |i_c\int_{-\mathcal{L}/2 }^{\mathcal{L}/2 }e^{ikx}dx\right |=i_c\left |\int_{-\mathcal{L}/2 }^{\mathcal{L}/2 }\cos(kx)dx\right |=\\
&=&\frac{I_c}{\mathcal{L}}\left |\int_{-\mathcal{L}/2 }^{\mathcal{L}/2 }\cos(kx)dx\right |,
\end{eqnarray}
and finally~\cite{SIBar82}
\begin{eqnarray}
\frac{I_s^m(\Phi )}{I_c}=\left | \frac{\sin \frac{\pi \Phi }{\Phi_0}}{\frac{\pi \Phi }{\Phi_0}} \right |,
\label{ShortJJDiffrPattern}
\end{eqnarray}
where $k=\frac{2\pi\mu_0t_d}{\Phi_0}H_{ext}$, $\Phi$ is the magnetic flux through the effective magnetic area ($t_d\,\mathcal{L}$), and $I_c=i_c \mathcal{L}=J_c\mathcal{L}\mathcal{W}=J_c\mathcal{A}$. 

The long junction case markedly differs with respect to the short case, since both the penetrating external field and the self-field generated by the Josephson current have to be considered, so that $\varphi(x)$ nonlinearly changes along the junction according to equations~(\ref{SGeq})-(\ref{bcSGeq}).
Therefore, in normalized units, the maximum value of the Josephson current can be written as~\cite{SIGia13,SIGua16}
\begin{equation}
\frac{I^m_s(t)}{I_c}= \frac{1}{L}\left |\int_{0}^{L} dx \cos \varphi(x,t)\right |.
\label{MaxNormHeatCurrent}
\end{equation}

It only remains to include in equation~(\ref{MaxNormHeatCurrent}) the proper phase difference $\varphi(x,t)$ for a driven LJJ given by solving equations~(\ref{SGeq})-(\ref{bcSGeq}). 
The magnetic field dependence of $I^m_s$ results in ``Fraunhofer-like'' diffraction patterns~\cite{SIKup06,SIKup10,SIGua16}.
While in the short junction limit~\cite{SIBar82}, different diffraction \emph{lobes} are well separated, here we observe the overlapping of the lobes.
The transitions between these lobes are usually discontinuous.
These patterns can be explained in terms of solitons entering the JJ. 

Each lobe corresponds to a state with a fixed number of solitons.
When the magnetic field increases, the configuration with more solitons is energetically favorable and, thus, the system jumps from a metastable state to a more stable state with more solitons.
In the region of $H$ values in which the diffraction lobes overlap, several solutions with different number of solitons may co-exist~\cite{SIKup06,SIKup10}.
Therefore, the system stays in the present configuration until the following one is energetically more stable.

To further explore the behavior of a magnetically driven LJJ, we have implemented a double-swept drive. The forward, i.e., with $H$ increasing, and the backward, i.e., with $H$ decreasing, patterns are significantly different.
For a given value of the magnetic field, the critical currents in the backward and forward evolutions differ and the system is found in a different diffraction lobe.
We can associate the forward and backward stable states (at fixed $H$) with a different number of solitons in the junction. Interestingly, the overall effect is a hysteric behavior in the critical current.

The diffraction patterns of the Josephson critical current, as the driving field is first increased (forward plot) and then reduced (backward plot), are shown in Fig.~\ref{FigSI02} for several JJ normalized lengths.

\section{The multistate structure as a function of the junction length}

\begin{figure*}[t!!]
\centering
\includegraphics[width=\textwidth]{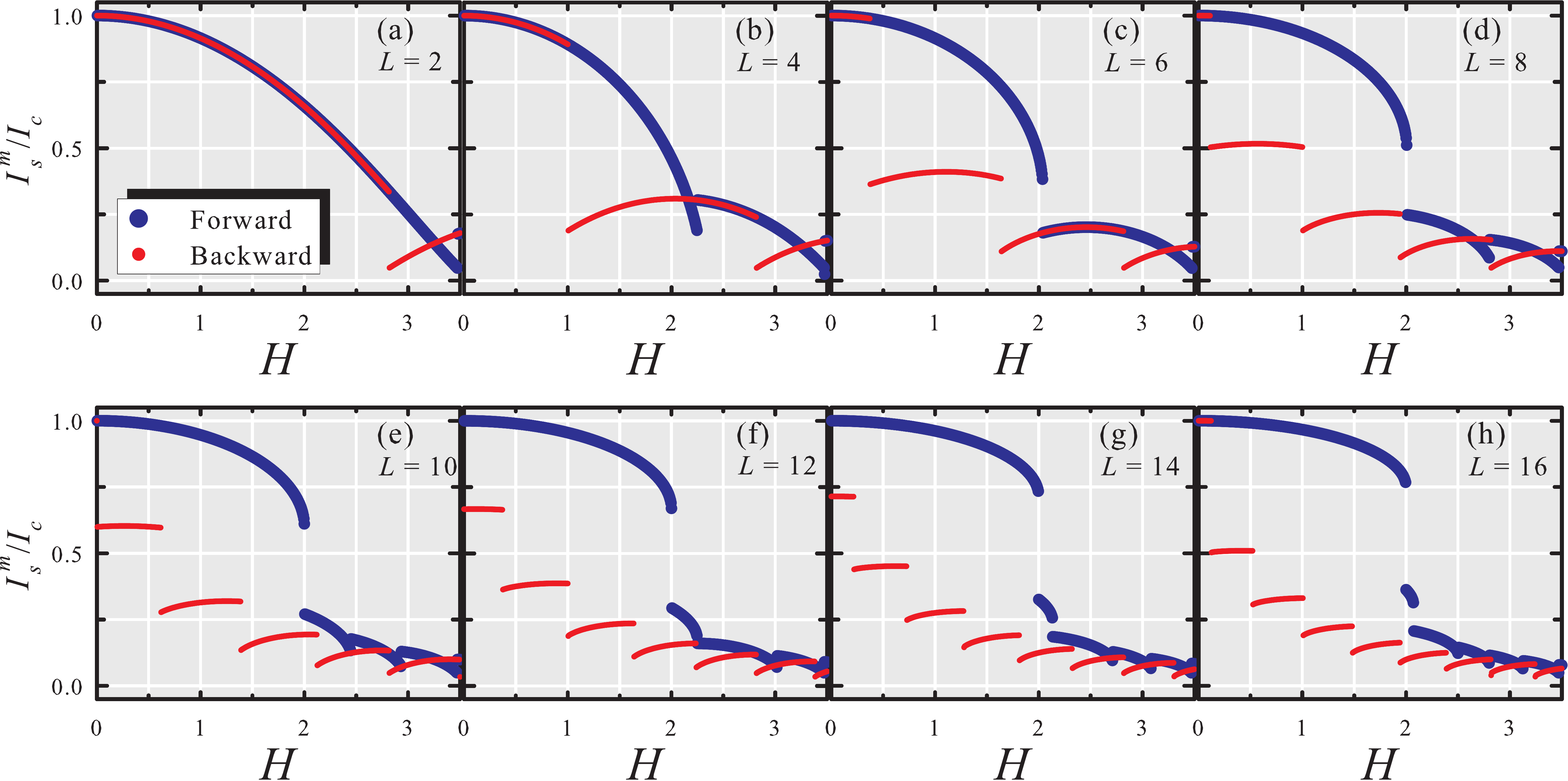}
\caption{
The forward and backward normalized critical current $I^m_s/I_c$ as a function of the magnetic field $H$ setting the damping parameter $\alpha = 0.24$ for several JJ length $L=2,4,6,8,10,12,14, \text{and } 16$ [panels (a), (b), (c), (d), (e), (f), (g), and (h) respectively]. Specifically, $H$ is swept first forward from $H\text{=}0$ and then backward. The legend in panel (a) refers to all panels.}
\label{FigSI02}
\end{figure*}

Forward-backward differences in the hysteretic behavior of the critical current are strongly evident for $|H|\lesssim H_c=2$, see Fig.~\ref{FigSI02}. 
In the forward pattern, the first lobe corresponds to the Meissner state, i.e., zero solitons in the junction, whereas by exceeding the threshold value $H_c$ the second lobe begins and solitons in the form of magnetic fluxons penetrate into the junction. This value of the critical field characterizes the diffraction patterns of the Josephson critical current in both overlap and inline LJJs~\cite{SIOwe67,SIBar82,SICir97}.
For $H>0$, the backward dynamics is described by $N$-solitons solutions, with $N\geq 1$. The amount of solitons exited depends on both the field intensity and the length of the junction. 
The many-soliton backward solutions suggest applications of this system as \emph{multi-state memories}, in which each state is clearly indicated by drastic suppressions of the critical current $I_s^{b}$ with respect to $I_s^{f}$. 
To quantify the LJJ memory-device performance we make use of a figure of merit defined by
the difference between the critical currents, 
\begin{equation}
\delta I_i=\frac{\left | I^f_s(H_i)-I^b_s(H_i) \right |}{I_c},
\label{IcDistances}
\end{equation}
where $H_i$ is the magnetic field at the midpoint of the $i$-th backward diffraction lobe.
For large $\delta I_i$ one can safely distinguish distinct memory states (MSs), namely, the current states. 
For instance, by focusing on the panels c,d, and e of Fig.~\ref{FigSI02}, we observe that, in the range $H\in [0-H_c]$, 
\begin{itemize}
\item for $L=6$, only one MS is clearly available, with a current difference $\delta I_1(H_1\simeq 1)\sim 0.5$, see Fig.~\ref{FigSI02}c;
\item for $L=8$, two MSs can be defined, with $\delta I_1(H_1\simeq 0.5)\sim 0.5$ and $\delta I_2(H_2\simeq 1.5)\sim 0.6$, see Fig.~\ref{FigSI02}d;
\item for $L=10$, three MSs can be defined, with $\delta I_{1}(H_1\simeq 0.32)\sim 0.4$, $\delta I_{2}(H_2\simeq 1)\sim 0.6$, and $\delta I_{3}(H_3\simeq 1.75)\sim 0.6$, see Figs.~\ref{FigSI02}e. 
\end{itemize}

Finally, junctions with different lengths are characterized by different numbers of distinct available MSs, each of them corresponding to a specific amount of solitons arranged along the junction.

Moreover, for a fixed effective junction length $\mathcal{L}$, the normalized length $L(T)=\mathcal{L}/\lambda_{_{J}}(T)$ and, therefore, the amount of MSs of the memdevice can be controlled by changing the temperature $T$ of the system (see below). 

\section{Physical quantities}

To give a realistic estimate of the physical quantities used in the computations, both the superconductors and the insulator making the junction (according to which distinctive values of resistance per area $R_a$ and specific capacitance $C_s$ of the junction result), and the normalized length of the device have to be chosen. Therefore, let us set a Nb/AlO/Nb junction, characterized by $R_a=50~\Omega~\mu\text{m}^2=5 \times 10^{-11}\Omega~ \text{m}^2$ and $C_s=50\frac{fF}{\mu m^2}= 5\times 10^{-2}\frac{F}{m^2}$, and a length-to-Josephson-penetration-depth ratio equal to $L=10$.\\
In the low temperature regime, the critical current is $I_c=\frac{\pi}{2}\frac{\Delta_0}{eR}=\frac{\pi}{2}\frac{\Delta_0}{eR_a}\mathcal{A}$ and, accordingly, $J_c=\frac{\pi}{2}\frac{\Delta_0}{eR_a}=\frac{\pi}{2}\frac{1.764k_bT_c}{eR_a}\sim0.44\times 10^8\text{A/m}^2=0.44\times 10^4\text{A/cm}^2$, $T_c=9.2\text{K}$ being the Nb critical temperature.

The main physical quantities to fully describe the system are:
\begin{itemize}
\item effective magnetic thickness \\$t_d=2\lambda^0_L+d\sim161\text{nm}$, the London penetration depth of a Nb thin film being $\lambda_{L}^0\sim80\text{nm}$ and setting $d=1\text{nm}$;
\item Josephson penetration depth \\ $\lambda_{_{J}}\text{=}\sqrt{\frac{\Phi_0}{2\pi \mu_0}\frac{1}{t_d J_c}}\sim 6\mu \text{m}$;
\item Linear dimensions \\$\mathcal{L}=10\lambda_{_{J}}=60\mu \text{m}\qquad$ and $\qquad\mathcal{W}=1\mu \text{m}$;
\item Area \\$\mathcal{A}=\mathcal{W}\mathcal{L}=60\mu \text{m}^2=6\;10^{-11}\text{m}^2$;
\item Critical current \\$I_c=J_c \times \mathcal{A}=0.44\;10^8 \text{A/m}^2\times 6\;10^{-11}\text{m}^2=2.67\times 10^{-3}\text{A}=2.67\text{mA}$;
\item Capacitance \\$C=C_s \times \mathcal{A}=5\;10^{-2}\times 6\;10^{-11}\text{F}=3\;10^{-12}\text{F}=3\text{pF}$;
\item Plasma frequency \\$\omega_p=\sqrt{\frac{2\pi}{\Phi_0}\frac{I_c}{C}}\sim1.63\textup{THz}$;
\item Resistance \\$R = \frac{R_a}{A}=\frac{5 \; 10^{-11}\Omega \textup{m}^2}{6\;10^{-11}\text{m}^2}\sim0.82\Omega$;
\item Damping parameter \\$\alpha =\frac{1}{\omega _p R C}\sim 0.24$;
\item Magnetic field \\$H=\frac{2\pi\mu _0}{\Phi _0}t_d\lambda_JH_{ext}$ so that 
$ H_{ext}=\frac{\Phi _0}{2\pi^2\mu _0}\frac{1}{t_d\lambda_{_{J}}}H\sim 3.4H\,\textup{Oe}$.
\end{itemize}

\section{Thermal effects}

\begin{figure*}[t!!]
\centering
\includegraphics[width=0.329\textwidth]{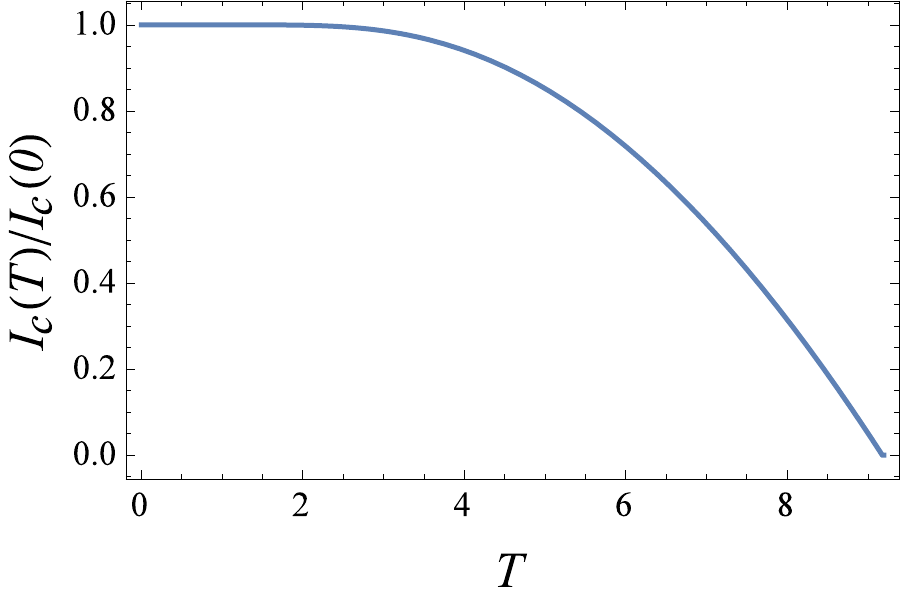}
\includegraphics[width=0.329\textwidth]{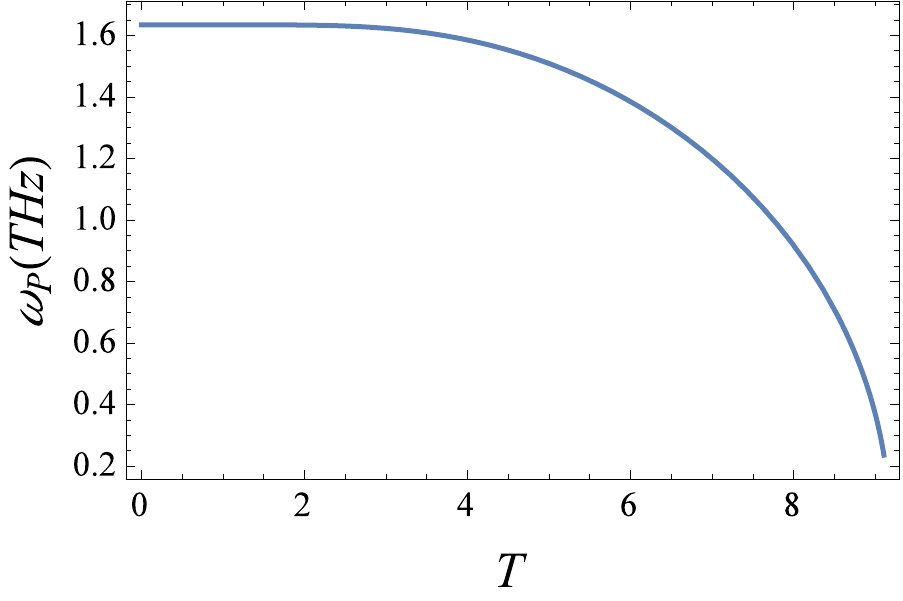}
\includegraphics[width=0.329\textwidth]{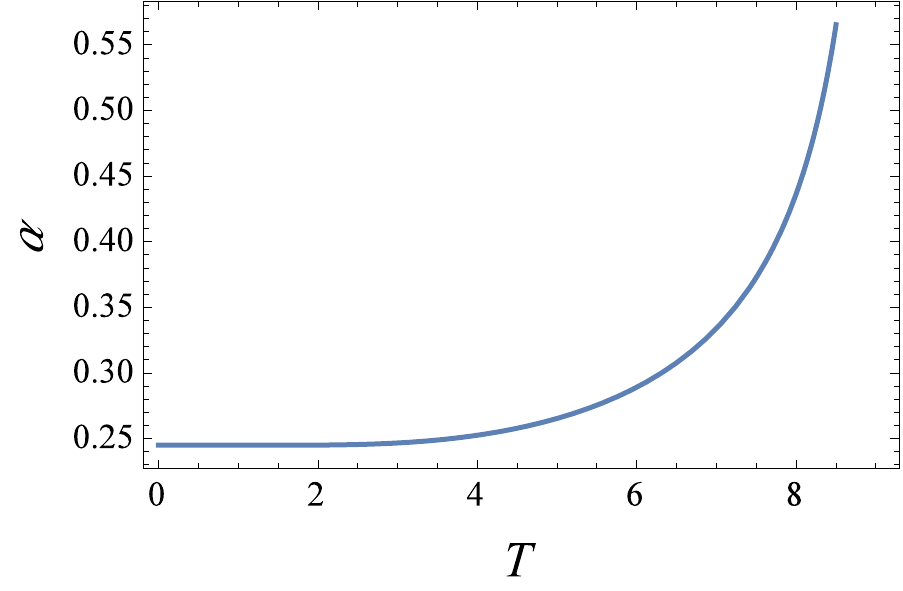}\\
\includegraphics[width=0.329\textwidth]{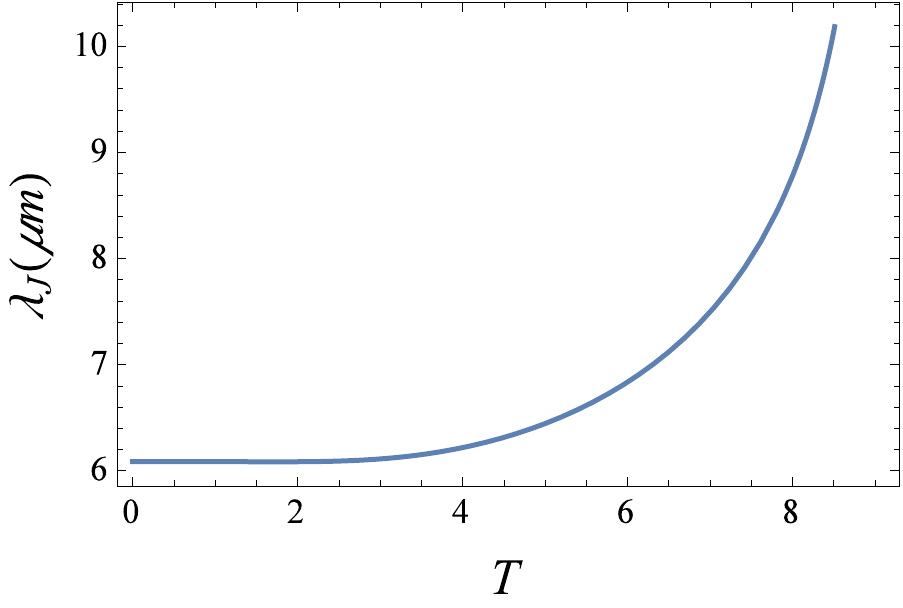}
\includegraphics[width=0.329\textwidth]{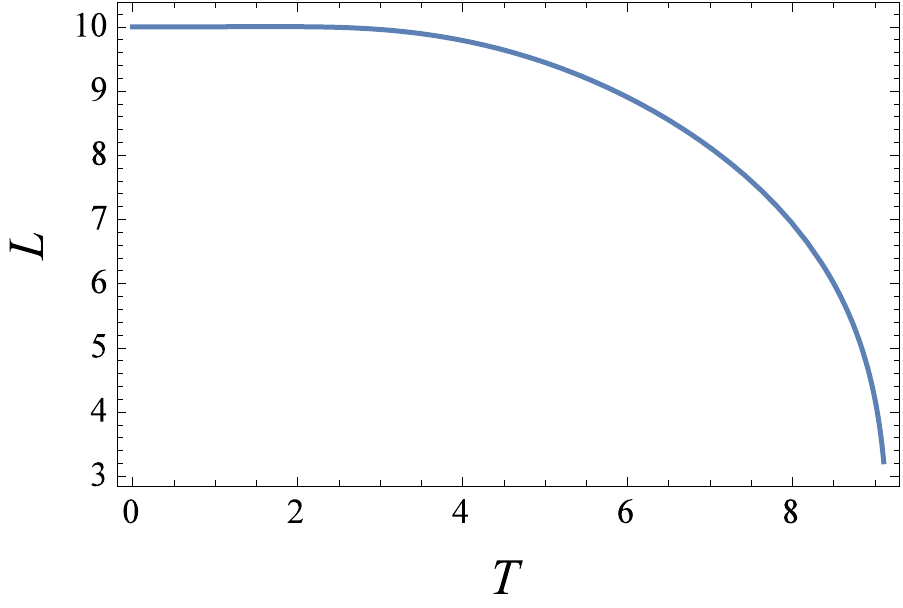}
\includegraphics[width=0.329\textwidth]{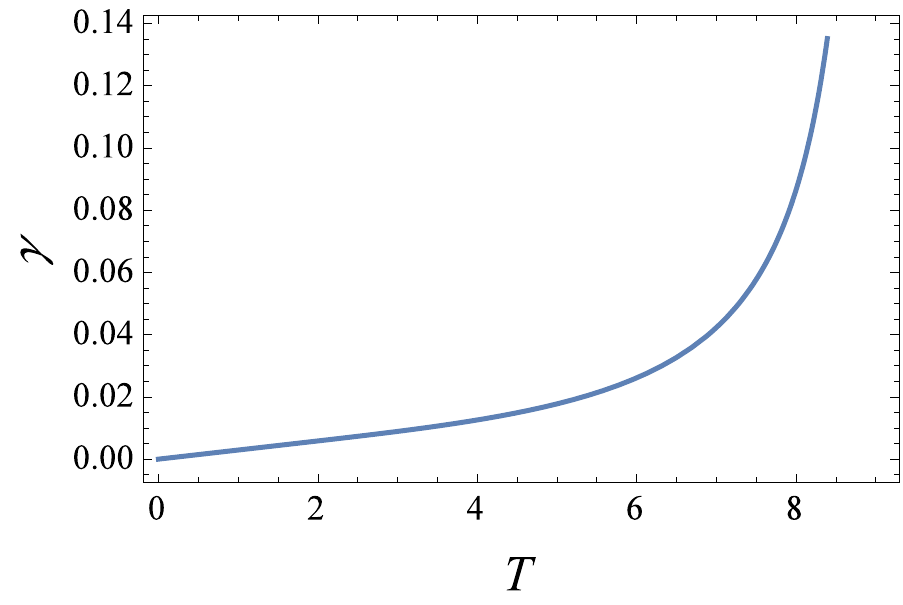}
\caption{Normalized Josephson critical current $I_c$, plasma frequency $\omega_p$, damping parameter $\alpha$, Josephson penetration depth $\lambda_J$, normalized length $L$, and thermal noise amplitude $\gamma$ as a function of the temperature, for $T_c=9.2K$.}
\label{FigSI03}
\end{figure*}
%

Some of the quantities introduced so far have an explicit dependence on the temperature. 
In particular, for identical superconductors,~\cite{SIBar82}
\begin{itemize}
\item the effective magnetic thickness $t_d(T)$ depends on $T$ through the London penetration depth $\lambda_i(T)=\lambda^0_{L}\Big/\sqrt{1-\left ( T_i/T_c \right )^4}$, and 
\item the Josephson critical current $I_c(T)$ depends on $T$ given by the Ambegaokar and Baratoff formula~\cite{SIBar82}
\begin{equation}\label{AmbBarFormula}
I_c(T)=\frac{\pi}{2}\frac{\Delta(T)}{eR}\tanh\left [\frac{\Delta(T)}{2k_bT}\right ],
\end{equation}
where $\Delta(T)$ is the BCS gap of the superconductors.
\end{itemize}
Accordingly, the plasma frequency $\omega_p(T)=\sqrt{\frac{2\pi}{\Phi_0}\frac{I_c(T)}{C}}$, the damping parameter $\alpha(T) =\frac{1}{\omega_p(T) R C }$, the Josephson penetration depth $\lambda_{_{J}}(T)=\sqrt{\frac{\Phi_0}{2\pi \mu_0}\frac{1}{t_d(T) J_c(T)}}$, and the normalized length $L(T)=\mathcal{L}/\lambda_{_{J}}(T)$ vary by changing the temperature.\\
\indent Concerning the normalized length, since $\frac{\lambda_{_{J}} (T)}{\lambda_{_{J}} (0)}=\frac{L (0)}{L (T)}$, if $L(T\to 0)=10$, as the temperature is increased to $T^*\sim 0.8T_c$, being $\lambda_{_{J}}(T^*)/\lambda_{_{J}}(0)\sim 1.3$, the corresponding normalized length becomes $L(T^*)\sim 8$.

\begin{table}[hb!!]
\centering
\begin{tabular}{|c|c|c|c|l}
\cline{1-4}
$T [\text{K}]$ & $\gamma(T)$ & $\alpha(T)$ & $\omega_p(T) [\text{THz}]$ & \\ \cline{1-4}
0.02 & 0.00006 & 0.244 & 1.634 & \\ \cline{1-4}
0.3 & 0.00088 & 0.244 & 1.634 &\\ \cline{1-4}
1.2 & 0.00351 & 0.244 & 1.634 &\\ \cline{1-4}
4.2 & 0.01346 & 0.254 & 1.573 &\\ \cline{1-4}
$0.75\times T_c=6.9$ & 0.03997 & 0.328 & 1.221 &\\ \cline{1-4}
$0.9\times T_c=8.3$ & 0.11660 & 0.494 & 0.810 &\\ \cline{1-4}
\end{tabular}
\caption{Thermal noise amplitudes, damping parameters, and plasma frequencies in correspondence of few specific temperatures used to obtain the results shown in the manuscript.}
\label{QuantitiesValues}
\end{table}

The temperature of the bath influences also the dynamics of the junction.
In order to take into account the thermal fluctuations on the phase dynamics, a noise current $i_T$ has to be included into the perturbed SG equation
\begin{equation}
\frac{\partial^2 \varphi }{\partial t^2}+\alpha(T)\frac{\partial \varphi }{\partial t}-\frac{\partial^2 \varphi }{\partial x^2} = - \sin(\varphi)+ i_T \left ({x,t}\right ).
\label{NoisySGeq}
\end{equation}
The normalized thermal current $i_T\left ({x,t}\right )$ is characterized by the well-known statistical properties of a Gaussian random process
\begin{eqnarray}
\left \langle i_T \left ({x,t}\right ) \right \rangle &=& 0 \\
 \left \langle i_T\left ({x,t}\right )i_T\left ({x',t'}\right ) \right \rangle &=& 2\gamma(T) \delta \left (x-x' \right )\delta \left (t-t' \right ).
\label{WNProperties}
\end{eqnarray}
For a LJJ, the thermal noise amplitude reads~\cite{SCas96}
\begin{equation}\label{NoiseAmplitude}
\gamma(T)=\frac{2\pi }{\Phi_0}L(T)\alpha(T)\frac{k_bT}{I_c(T)}.
\end{equation}
The behaviors of the normalized Josephson critical current $I_c$, the plasma frequency $\omega_p$, the damping parameter $\alpha$, the Josephson penetration depth $\lambda_J$, the normalized length $L$, and the thermal noise amplitude $\gamma$ as a function of the temperature, for $T_c=9.2K$, are shown in Fig.~\ref{FigSI03}. 

Moreover, the values of the thermal noise amplitude, the damping parameter, and the plasma frequency in correspondence of few specific temperatures (used to obtain the results shown in the manuscript) are listed in Table~\ref{QuantitiesValues}.

\section{The frequency response}

\begin{figure*}[htpb!!]
\centering
 \includegraphics[width=\textwidth]{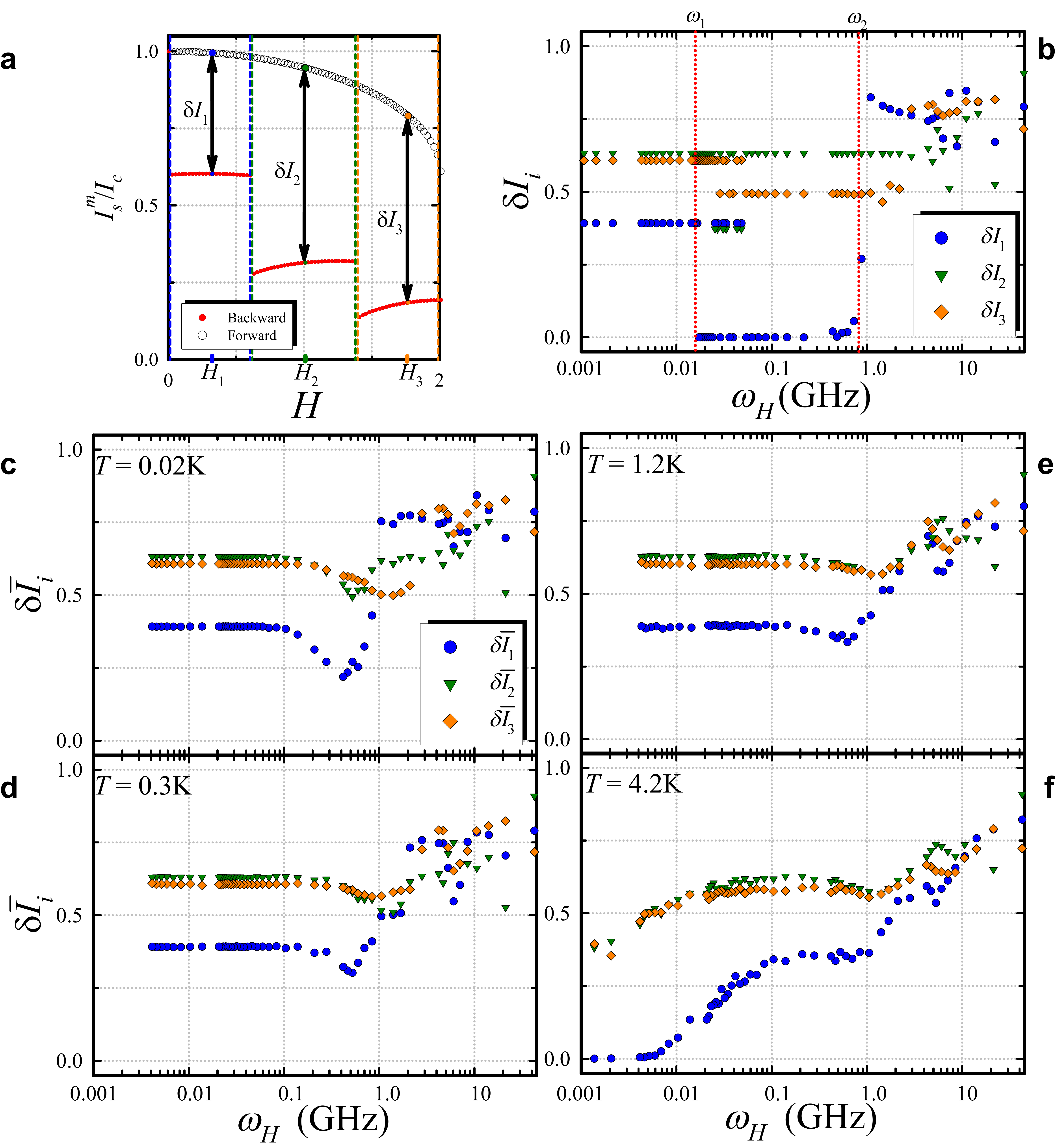}
\caption{
\textbf{a}, Forward and backward diffraction patterns for $H\in[0-2]$ and $L=10$. The backward pattern is composed by three lobes in the place of the large lobe of the forward one. The value of the magnetic field $H_i$ in the center of each lobe and the differences $\delta I_i$ ($i=1,2,3$), see equation~(\ref{IcDistances}), for $H\equiv H_i$ are also shown. The values of $\delta I_i$ are used to check the behavior of the logic states of the LJJ-based memory against frequency variations. \textbf{b}, $\delta I_i$ ($i=1,2,3$) as a function of the driving frequency $\omega_H=\omega_p/T_H$ in absence of thermal noise. As $\omega_H$ reduces, the diffraction patterns tend to become steady and $\delta I_i$ approach constant values. Specifically, by defining two threshold values, $\omega_1$ and $\omega_2$, the behavior of the device in different ranges of frequencies can be discussed: \emph{i}) for $\omega_H\gtrsim \omega_2$ the system is not able to respond to extremely high driving frequency oscillations; \emph{ii}) in the range $\omega_H\in[\omega_1-\omega_2]$ the memory cannot safely provide three logic states ; \emph{iii}) for $\omega_H\lesssim\omega_1$ the distances $\delta I_i$ approach constant values, and, in spite of frequency variations, the system provides three distinct states. \textbf{c}, \textbf{d}, \textbf{e}, and \textbf{f}, distances $\delta \overline{I_i}$ ($i=1,2,3$), see equation~(\ref{MeanIcDistances}), as a function of $\omega_H$ for $T=\{0.02, 0.3,1.2, 4.2\} \text{K}$, respectively. The legend in panel \textbf{c} refers to these panels too.}
\label{FigSI04}
\end{figure*}

Independently of the physical mechanism defining the state of the device, the memelement response is usually strongly dependent of the frequency of the input drive~\cite{SIPerDiV11}. At low frequencies, the system has enough time to adjust its state to the instant value of the drive, so that the device non-linearly behaves and a hysteretic evolution results. Conversely, at high frequencies, there is not enough time for any change during an oscillation period of the drive. Therefore, we explore the effects of variations of the driving frequency on the behavior of our device. However, our system is an example of a memory that benefits from, and properly work only in, the presence of noise. To discuss this point, we compare results obtained in both the deterministic and stochastic approaches, by taking into account several temperatures. 

Recently, the effects of the noise on the performance of several memory devices has been investigated~\cite{SISto12,SISli13,SIPat13,SIPatFie13,SIPat14,SIFie14,SIPat15,SIPat16}.

To quantify the LJJ-based memory performances as the driving frequency is changed, we use the distances $\delta I_i$ defined in equation~(\ref{IcDistances}). Specifically, Fig.~\ref{FigSI04}a shows the midpoint values $H_i$ of the backward diffraction lobes within the field range $H\in[0-2]$, and the distances $\delta I_i$ ($i=1,2,3$) for these fields, for a junction with $L=10$. 

First, we analyse the device performance in absence of thermal noise.
The behavior of $\delta I_i$ ($i=1,2,3$) as the driving frequency $\omega_H$ is changed is shown in Fig.~\ref{FigSI04}b for the deterministic case, i.e., no noise source is considered in the model. For the sake of clarity, we define in Fig.~\ref{FigSI04}b two threshold frequencies, $\omega_1$ and $\omega_2$, and examine the results in different frequency ranges:
\begin{itemize}
\item for $\omega_H\lesssim\omega_1$ the distances $\delta I_i$ approach constant values, inasmuch steady diffraction patterns are obtained. For these frequencies, the logic states are definitively robust against frequency variations;
\item in the range $\omega_H\in[\omega_1-\omega_2]$ the values of $\delta I_i$, and accordingly the amount of the logic states, significantly deviate from the steady ones; 
\item for $\omega_H\gtrsim \omega_2$ the system is not able to respond to extremely high driving oscillations, so that the backward patterns are highly disordered, despite the fact that $\delta I_i\to1$, and therefore the logical states cannot be safely distinguished. 
\end{itemize}

As discussed above, realistic devices are subject to thermal noise.
Far away from the critical temperature, the addition of the thermal noise has the effect to stabilise the dynamics and, therefore, to access to higher driving frequencies.
\begin{figure}[htbp!!]
 \includegraphics[width=0.5\textwidth]{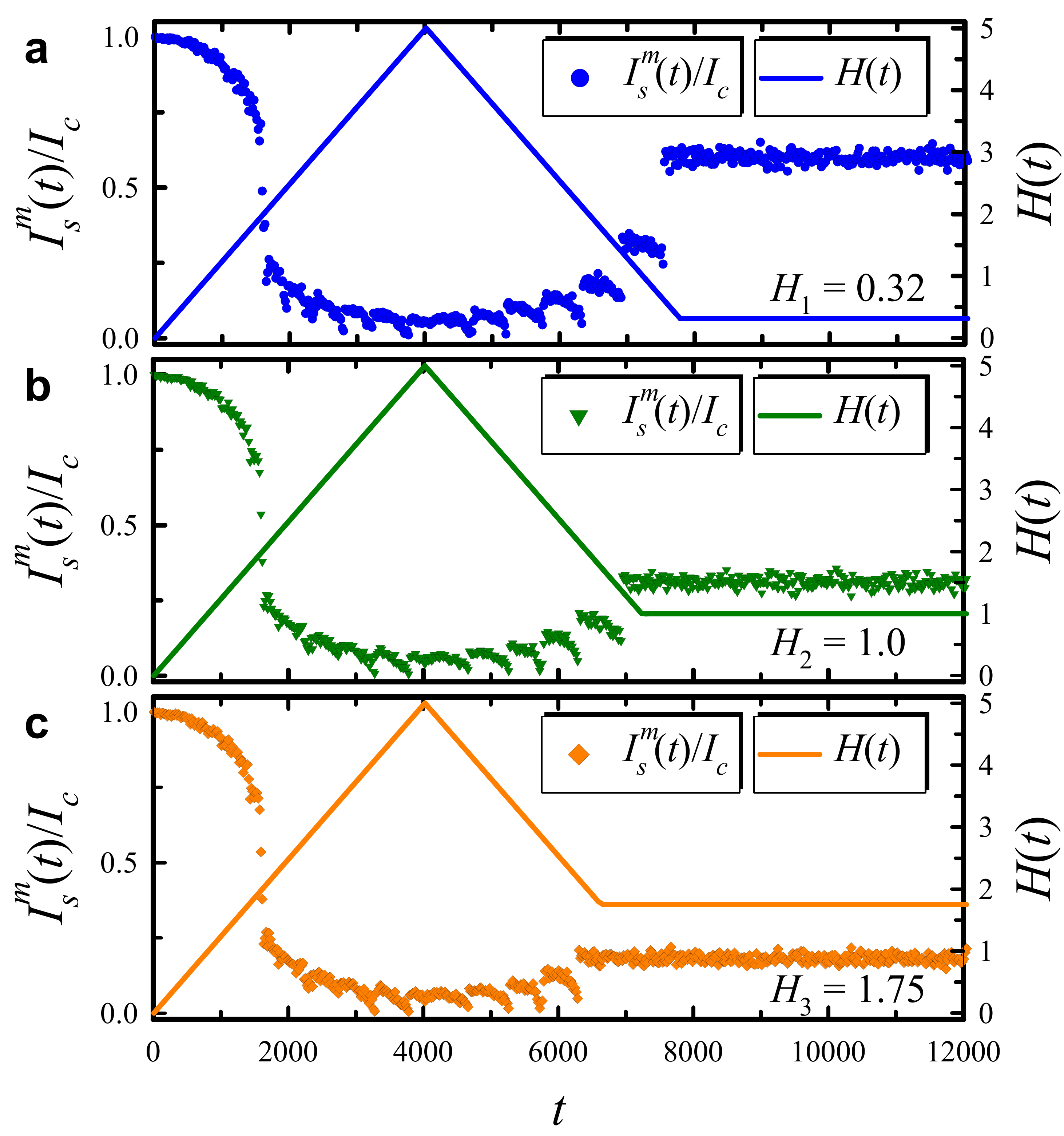}
\caption{Normalized critical current $I_s^m/I_c$ (left ordinate scale, full symbols) and driving field $H$ (right ordinate scale, solid lines) as a function of the time $t$, normalized to $\omega_p$, for the MSs defined in Fig.~\ref{FigSI04}a for $L=10$ and $H_i=0.32, 1.0, 1.75$ with $i=1,2,3$ (panels a, b, and c, respectively). The magnetic field value $H_i$ is chosen in the midpoint of the $i$-th backward diffraction lobe, so that the robustness against small field fluctuations is ensured. These graphs are obtained by setting $H(t\geq t_i)=H_i$, where $t_i$ is the time for the magnetic field $H(t)$ to reach the value $H_i$ during the backward sweep. The diffraction patterns are computed for $T=1.2K$. In spite of the thermal fluctuations, as the magnetic field is set to $H(t)=H_i$, the $\delta I_i$ values are practically constant in time, i.e., $I^m_s(t\geq t_i)\sim I^m_s(t_i)$. These results are obtained by setting $\Delta t_H=4$, $\Delta H=0.005$, and $H_{max}=5$, so that the non-normalized driving frequency is $\omega_H=\omega_p/T_H\simeq 0.1\textup{GHz}$, where $\omega_p(T=1.2\textup{K})\simeq1.634\textup{THz}$ (see Table~\ref{QuantitiesValues}). }
\label{FigSI05}
\end{figure}

In the noisy approach, the distances 
\begin{equation}
\delta \overline{I_i}=\frac{\left | \overline{I^f_s}(H_i)-\overline{I^b_s}(H_i) \right |}{I_c},
\label{MeanIcDistances}
\end{equation}
are taken into account. The behaviors of $\delta \overline{I_i}$ ($i=1,2,3$) as a function of $\omega_H$ for $T=\{0.02, 0.3,1.2, 4.2\} \text{K}$ are shown in panels c, d, e, and f of Fig.~\ref{FigSI04}, respectively. The quantities $\overline{I^f_s}/I_c$ and $\overline{I^b_s}/I_c$ are computed by averaging over the total number of numerical realizations, $N_{exp}=100$, the normalized critical currents as the magnetic field $H$ is swept forward and backward, respectively, when the thermal fluctuations are included in the SG model.

For $T=0.02\textup{K}, 0.3\textup{K}, \textup{and } 1.2\textup{K}$ (see Figs.~\ref{FigSI04}c, d, and e, respectively) the values of $\delta \overline{I_i}$ are roughly constant and the logic states of the device are definitively stable up to $\omega_H\sim 0.1 \textup{GHz}$. Conversely, for higher frequencies, the inability of the system to adjust its state to rapid changes in the magnetic bias comes to light. 

For $T=4.2\textup{K}$, i.e., the liquid helium temperature, the frequency behavior significantly changes, see Fig.~\ref{FigSI04}f. In fact, $\delta \overline{I_i}$ ($i=1,2,3$) approach the values obtained for low temperatures only for $\omega_H\sim0.1\textup{GHz}$. Conversely, for lower frequencies the thermal fluctuations have enough time to guide the evolution of the system, so that the state of the system is set by noise-induced transitions. Therefore, the backward and forward patterns tend to superimpose and $\delta \overline{I_i}\to0$.

Moreover, we verify if the system is able to provide information-storage times longer than any practical reading times, so that it works as a \emph{non-volatile memory}~\cite{SIDiVPer13}. To this end, we show in Fig.~\ref{FigSI05} the normalized critical currents $I_s^m/I_c$ and driving field $H$ as a function of the normalized time $t$ for the states defined in Fig.~\ref{FigSI04}a, for $T=1.2K$. Specifically, results in panels a, b, and c of Fig.~\ref{FigSI05} are obtained by freezing the magnetic field to $H(t\geq t_i)=H_i$ with $H_i=0.32, 1.0, \textup{and } 1.75$, respectively, $t_i$ being the time for the magnetic field to reach the value $H_i$ during the backward sweep. In spite of the thermal fluctuations, as the magnetic field is set to $H(t)=H_i$, the critical current is roughly constant, i.e., $I^m_s(t\geq t_i)\sim I^m_s(t_i)$, so that steady logic states are established.

\section{Memory devices and the response function}

The properties of a memory-element (\emph{memelement} for short) depend on the state and the history of the system~\cite{SIPer11}. Specifically, in \emph{ideal memristors}, \emph{memcapacitors} and \emph{meminductors}~\cite{SIPer11}
\begin{itemize}
\item (ideal memristor) the resistance depends only on the charge that flows in the system (or on the history of the voltage);
\item (ideal memcapacitor) the capacitance depends only on the history of the charge stored on its plates (or the history of the voltage across it);
\item (ideal meminductor) the inductance only depends on the history of the current that flows through it (or the history of the flux).
\end{itemize}
These definitions can be generalized by invoking a general non-linear, memory-dependent response function $g$~\cite{SIDiV09}
\begin{eqnarray}\nonumber
y(t)&=&g(x,u,t)u(t)\\
\dot{x}&=&f(x,u,t)
\label{MemRelations}
\end{eqnarray}
where
\begin{itemize}
\item $g(x,u,t)$ is the response function;
\item $u(t)$ is the input signal;
\item $y(t)$ is the output signal;
\item $f(x,u,t)$ vector function of internal state variables;
\item $x$ vector of internal state variables.
\end{itemize}
Generally, in real systems ideal memdevices are usually rare, so that the relation between current and voltage defines a \emph{memristive system} (i.e., the resistance depends on both the charge and other internal variables of the system), while the relation between charge and voltage specifies a \emph{memcapacitive system}, and the flux-current relation gives rise to a \emph{meminductive system}~\cite{SIPer11}.

A distinctive signature of memory devices is the presence of a \emph{hysteresis loop} in the behavior of the output $y(t)$ and/or the response function $g(t)$ as a function of the input $u(t)$~\cite{SIPer11}. The features of the hysteresis loop depend on the properties of both the system and the input $u(t)$, such as its amplitude and frequency. Hysteresis loops can be \emph{pinched}, when the loop passes through the origin ($y$ is zero whenever $u$ is zero and vice versa). Moreover, a pinched hysteresis can be self-crossing~\cite{SIPer11}, i.e., with the crossing between opposite direction branches of the loop, or not self-crossing.

In contrast with usual memelements defined by equations~(\ref{MemRelations}), the behavior of our LJJ-based memory-device is not directly stated in the form of a relation between $I^m_s(t)$ and $H(t)$ through a response function $g(\varphi,H,t)$. In other words, our memdevice is not described by a current-field expression such as $I^m_s(t)=g(\varphi,H,t)H(t)$.
In fact, in normalized units, the critical current reads

\begin{equation}
\frac{I^m_s(t)}{I_c}= \frac{1}{L}\left |\int_{0}^{L} dx \cos \varphi(x,t)\right |.
\label{MaxNormHeatCurrent1}
\end{equation}

The internal state variable of this field-controlled memelement is the phase difference $\varphi(x,t)$, whose dynamics is ruled by equations~(\ref{SGeq})-(\ref{bcSGeq}). 

We observe that a relation including a response functional comes to light by first-order expanding the $\cos\varphi(x,t)$ term in equation~(\ref{MaxNormHeatCurrent1}) around the junction edge $x=0$, that is by ignoring the non-linearity of the problem,
\begin{equation}
\cos\varphi(x,t)\sim_{_{x=0}}\cos\varphi(0,t)-\sin\varphi(0,t)\left .\frac{\mathrm{d} \varphi(x,t)}{\mathrm{d} x} \right |_{_{0}}x.
\end{equation}
Therefore, equation~(\ref{MaxNormHeatCurrent1}) becomes
\begin{eqnarray}
&&\frac{I^m_s(t)}{I_c}=\left | \frac{1}{L}\int_{0}^{L}\cos\varphi(x,t)dx \right |\sim\\ \nonumber
&&\sim \left | \frac{1}{L}\int_{0}^{L}\cos\varphi(0,t)dx -\frac{1}{L}\int_{0}^{L}\left [ \sin\varphi(0,t)\left .\frac{\mathrm{d} \varphi(x,t)}{\mathrm{d} x} \right |_{_{0}}x \right ]dx \right |.
\end{eqnarray}
According to equation~(\ref{bcSGeq}), the previous equation reads
\begin{eqnarray}\nonumber
\frac{I^m_s(t)}{I_c}&\sim& \left | \frac{1}{L}\cos\varphi(0,t)L -\frac{\sin\varphi(0,t)}{L}H(t)\int_{0}^{L}xdx \right |\sim \\
&\sim&\left |\cos\varphi(0,t)+ F(\varphi,H,t)H(t) \right |.
\label{TaylorExp}
\end{eqnarray}
Here, we have defined the functional (i.e. response functional) $F(\varphi,H,t)$
\begin{equation}
F(\varphi,H,t)=-\sin\varphi_{_H}(0,t)\frac{L}{2},
\end{equation}
where the field-dependence of the phase dynamics is stressed.

\bibliography{biblio.bib}

\end{document}